# Quantized conductance observed during sintering of silver nanoparticles by intense terahertz pulses

Keisuke Takano,[1,2]* Hirofumi Harada,[2] Masashi Yoshimura,[2] and Makoto Nakajima[2]

[1]*Center for Energy and Environmental Science, Shinshu University, 3–1–1 Asahi, Matsumoto, Nagano 390–8621, Japan.*
[2]*Institute of Laser Engineering, Osaka University, 2–6 Yamadaoka, Suita, Osaka 565–0871, Japan*

*Corresponding author: ksk_takano@shinshu-u.ac.jp

**Abstract**

We show that silver nanoparticles, which are deposited on a terahertz-receiving antenna, can be sintered by intense terahertz pulse irradiation. The conductance of the silver nanoparticles between the antenna electrodes is measured under the terahertz pulse irradiation. The dispersant materials surrounding the nanoparticles are peeled off, and conduction paths are created. We reveal that, during sintering, quantum point contacts are formed, leading to quantized conductance between the electrodes with the conductance quantum, which reflects the formation of atomically thin wires. The terahertz electric pulses are sufficiently intense to activate electromigration, i.e., transfer of kinetic energy from the electrons to the silver atoms. The silver atoms move and atomically thin wires form under the intense terahertz pulse irradiation. These findings may inspire nanoscale structural processing by terahertz pulse irradiation.

With the advancement of wavelength conversion techniques, table-top intense terahertz pulse sources whose peak electric field is over 1 MV/cm have been developed.[1,2] Terahertz waves can be employed not only for linear spectroscopic analysis but also to control material properties. Nonlinear phenomena such as the phase change of $VO_2$,[3] luminescence,[4,5] carrier multiplication through impact ionization,[6] laser spectral broadening,[7] and electron emission [8–10] have been demonstrated using intense terahertz pulses.[11] The too much intense electromagnetic field can affect the material structure[12,13] and induce irreversible property changes. However, irreversible changes can be utilized in material processing such as powder sintering and ablation using microwave oven and laser irradiation.[14] It has been reported that irreversible morphological change is induced by intense terahertz field.[15] Seo *et al*, has been reported the ionization and accumulation of carbon atoms are induced in the vicinity of nano-antenna by enhanced terahertz pulses.[16] In addition, the investigation of the device damage caused by intense terahertz waves is important for terahertz applications.

In this study, we demonstrate the sintering of silver nanoparticles by intense terahertz pulses. We revealed that the conductance is quantized during the sintering process performed under atmospheric pressure and room temperature. The silver atoms move, and nanowires with quantum point contacts form owing to the intense terahertz electric pulses. The structural control of nanowires and nanogaps has been investigated for applications in sensors and molecular electronics.[17–19] Nanogaps can be fabricated through electromigration by applying DC voltage to nanowires. The results obtained in this study demonstrate the possibility of fine structural control of nanowires and nanogaps by low-photon-energy ultrafast terahertz electromagnetic pulses.

The change in conductance of the silver nanoparticles under the irradiation with intense terahertz pulses was investigated. The incident terahertz pulses were generated from regenerative amplified Ti:sapphire laser pulses using the tilted pulse-front method through a lithium niobate ($LiNbO_3$) prism.[1,2] Figure 1(a) shows the time-domain waveform and Fourier-transform spectrum of the terahertz pulse at the focal point. The terahertz pulses were focused on the samples using off-axis parabolic mirrors. The terahertz pulses were focused to a diameter of approximately 500 μm. The electric fields were measured using the electro-optical sampling method with a 1-mm-thick GaP (110) crystal. The terahertz pulses were almost mono-cycle pulses with a period of 2 ps and peak amplitude of 160 kV/cm. The repetition rate of the pulses was 1 kHz.

Figure 1(b) shows a schematic of the experiment. An aluminum bow-tie antenna was fabricated on a high-resistivity silicon substrate using the photolithography method. There is a gap with a width of 2 μm at the center of the triangular electrode pair. The thickness of the antenna was approximately 200 nm. The terahertz waves were focused at the gap; the fields were enhanced between the gap. Silver-nanoparticle dispersion liquid (Ag1teH, nanometal ink, Ulvac, Inc.) was introduced at the electrode gap using the inkjet printing method, [20,21] as shown in the schematic in Fig. 1(b). The nanometal-ink is utilized as an ink in printed electronics and electric joining materials. The nanoparticles, whose diameter is approximately 5 nm, are surrounded by the dispersant material and colloidally dispersed in the tetradecane solvent. A bias voltage of 1 mV was applied to the electrode pair. The current was measured (Keithley 2400 SourceMeter) under the irradiation with the terahertz pulses. The polarization of the terahertz pulses was parallel to the bias voltage.

Figures 2(a) and 2(b) show typical micrographs of the sample, before and after the terahertz irradiation, respectively. Before the terahertz irradiation, the silver nanoparticles are insulated by the dispersant. After an 8-h-long irradiation with the terahertz pulses, a color change to black was observed at the gap. The scanning electron microscopy image in Fig. 2(c) shows the presence of cracks. In the blacked cracked area, the nanoparticles form aggregates with a diameter larger than 100 nm (see Fig. S1-3 in the Supplemental Material). The terahertz irradiation peeled off the dispersant material and cracks were created with the decrease of the ink volume. The nanoparticle aggregates create conduction paths between the antenna gap. In order to peel off the dispersant, typically annealing at a temperature larger than 200 °C is required.[21] From these experimental facts, therefore, we conclude that the temporally and spatially localized heating induced by terahertz fields decomposes the dispersant. Typically, several hours of irradiation are required as the process is slow. It is worth noting that the color change and creation of the conduction path are not observed when the irradiation is performed using near-infrared laser pulses with the same repetition rate (1 kHz), average power (~ 1 mW), and beam diameter (see Fig. S4 in the Supplemental Material).

Figure 3(a) shows the currents through the gap of two same shape samples under terahertz irradiation. The currents were measured every 10 s for 8 h. First, the nanoparticles are insulated. Then, the currents increase during the repetitive creation and breaking of the conduction paths. The current changes infrequently and discontinuously. For the sample 1, the current increases abruptly after approximately 2 h of irradiation. Once the dispersant materials surrounding the nanoparticles are removed, neighboring nanoparticles start to aggregate. The aggregation gradually proceeds between the antenna gap. When the gap is connected with nanoparticles without dispersant, the current increases. After approximately 8 min, the connection broke and the current decreased. The thin wires can be easily broken. Then, the current increased again with discontinuous changes. Figure 3(b) shows an enlarged plot of the conductance change between 2.5 h and 3.5 h. The conductance is normalized to the conductance quantum $G_0 = 2e^2/h \approx 7.75 \times 10^{-5}$ S, where, $e$ and $h$ are the elementary charge and Planck constant, respectively. We observed that the conductance change is quantized with the conductance quantum. The Landauer formula describes the conductance of quantum point contacts as $G_n = nG_0$,[22] where the integer $n$ is the channel number of the electron transport. For the sample 1, conductance with $n = 1, 3,$ and $4$ is observed in the considered time range. The quantization of the conductance reflects the formation of atomically thin wires. For $n = 1$, the smallest cross-section of the conduction path is expected to be a single silver atom.[19] As far as the atomically thin wire is maintained, the electron wave function in the wire is restricted and the conductance is quantized even at room temperature. The conductance with $n = 1$ was stable for 15 min and then increased to values corresponding to $n = 3$ and 4. The trend of the conductance change depends on the sample. For the sample 2, conductance with $n = 1$ and 2 was observed, as shown in Fig. 3(c). However, it is worth noting that the parallel connection of the two quantum point contacts at $n = 1$ corresponds to a conductance of $G = 2G_0$. In this experiment, we cannot distinguish the increase of $n$ and creation of parallel connections of the quantum point contacts. However, the results show that atomically thin conduction paths can be formed by intense terahertz pulse irradiation.

We revealed that the silver nanoparticles are sintered, and atomically thin wires can be formed under the irradiation with the intense terahertz pulses. Previous studies have reported ionization and electron emission from metals irradiated by intense terahertz pulses.[8,9,13,23,24] The emitted electrons are accelerated by the electric field; their kinetic energy reaches several hundred electron volts along the antenna gap. Strikwerda *et al.* and Zhang *et al.* reported the deformation of the Au-antenna shape under the irradiation with intense terahertz pulses.[12,13] The emitted electrons are accelerated along the antenna gap (several micrometers) and collide the other side of the antenna electrode. The impact heating caused by the electron bombardment is sufficient to melt the Au electrodes, as observed in their study. However, in our experiment, the antenna gap contained silver nanoparticles, dispersant materials, and tetradecane. In such conditions, the electron heating by tunneling currents is a feasible origin of the decomposition of the dispersant materials. Liu *et al.* demonstrated an insulator-to-metal transition of vanadium dioxide ($VO_2$) on planar metallic resonators.[3] The terahertz fields that are enhanced in the resonator gap induced ionization and increase of the conductance of $VO_2$. They estimated that the electron temperature can rise approximately 100 K in $VO_2$ with a terahertz pulse enhanced by the antenna electrodes. The peak amplitude of the terahertz pulse was estimated to be 1 MV/cm in the vicinity of the antenna. Figure 4 shows the electric field along the gap of the bowtie antenna on a Si substrate simulated using the finite-difference time-domain method (Poynting for Optics, Fujitsu). In the simulation, the electrodes and the gap area are assumed to be a perfect electric conductor and air, respectively. The peak amplitude of the enhanced terahertz field is estimated to be 6 MV/cm at the vicinity of the electrodes. This field exceeds the damage threshold of $VO_2$ observed in Ref. 3. Although the situation in terms of materials is more complex in our experiment, which requires to investigate the thermal parameters of each material, the local temperature increase of approximately 200 K to resolve the dispersant material is presumable by the enhanced terahertz pulses. The resolving process of the dispersant material proceeds locally, and the gap suddenly becomes conductive when it is percolated with the uncovered nanoparticles. The conduction paths provided by the silver nanoparticles are unstable, and repeatedly connect and break under the terahertz pulse irradiation, as shown in Fig. 3(a).

Another feature observed in our experiment is the quantization of the conductance. It reflects the formation of atomically thin wires by the terahertz irradiation. One of the possible mechanisms of atomic movement is electromigration through kinetic energy transfer from electrons to atoms. Umeno *et al*. demonstrated the fabrication of a nanogap in an Au-nanowire by electromigration. [17,25] When the kinetic energy of the electrons is larger than the surface diffusion potential of the atoms (0.1–0.4 eV for Au surface), the atoms start moving. They revealed that when the voltage applied to the wire exceeds 0.1 V, the atoms start moving, which leads to changes in the conductance of the wire. The diffusion potential of silver atom is 0.1–0.4 eV, depending on the crystal face.[26] It is noted that this energy simply corresponds to the temperature of 1000–4600 K, which is much larger than room temperature. The DC voltage applied to measure currents in our experiment was 1 mV, significantly lower than the activation voltage of the electromigration. The local electric field applied by the terahertz pulses of 1–6 MV/cm corresponds to 0.1–0.6 V/nm, which implies that the electron kinetic energy becomes 0.1–0.6 eV after 1 nm of accelerated movement. The silver atoms can move owing to the kinetic energy received from the electrons. Once few atoms accidentally provide a conduction path with the quantum point contact, the electron transport there becomes ballistic. Therefore, the quantum point contacts are maintained and the atoms around the quantum point contact move more easily and make the contact thicker. If the acceleration distance is several micrometers, the kinetic energy of the electrons reaches several hundred electron volts. [12] Such kinetic energy could cause a vigorous evaporation of atoms. However, during the sintering of the silver nanoink, a large number of nanogaps spontaneously emerge. In the nanogaps, quantum point contacts can form owing to the atomic movement under the terahertz irradiation.

In summary, nonlinear phenomena induced by intense terahertz fields have been reported since the development of intense terahertz sources. In contrast to near-infrared laser irradiation, tunneling ionization and electron acceleration are remarkably induced because of the longer wavelength. In contrast to DC electronics techniques, electrodeless ultrafast electron emission and acceleration can be induced by the terahertz pulses. In this study, we showed that silver nanoparticles can be sintered by intense terahertz pulse irradiation. During the sintering process, quantized conductance was observed. A movement of silver atoms and formation of quantum point contacts were observed under the terahertz pulse irradiation. The electromigration caused by the terahertz electric field pulses is a feasible mechanism for the formation of the atomically thin wires. In a sense, the stepped conductance memorizes the currents induced by terahertz fields as a resistive memory although the change is irreversible and still not controllable in this study. At the same time, the quantized step of the conductance is the minimal predictor of the device damage. To fully understand the formation of the quantum point contacts induced by terahertz pulses, the investigation under the well-designed systems such as a single nanogap with different materials in vacuum is required to avoid the effect from the randomness of the structure and the materials other than the metals. However, these phenomena shown here may inspire not only a structural processing of nanostructures but also quantum state control and device reliability evaluation by intense terahertz pulses.

**Supplementary Material**

See supplementary materials for additional experiments of sintering silver nanoparticles under intense terahertz pulse irradiation.

**Acknowledgment**

The authors thank Dr. K. Murata and Dr. Y. Kashiwagi for their fruitful discussions. This study was partially supported by the Japan Society for the Promotion of Science (JSPS) KAKENHI (16H06025).

**Figures**

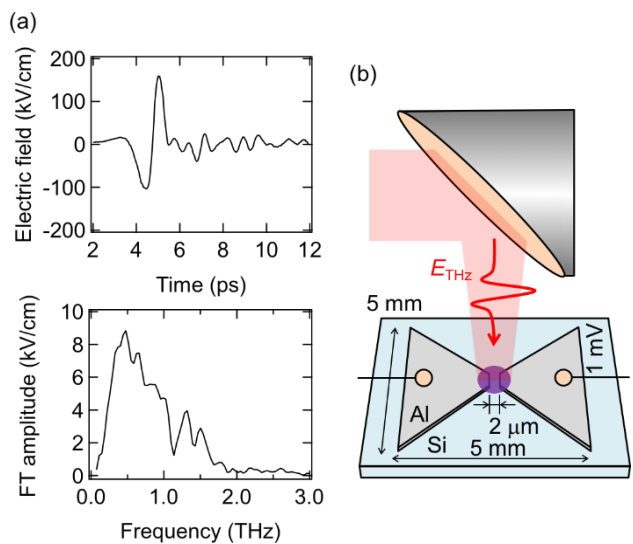

FIG. 1. (a) Time-domain waveform and Fourier-transform amplitude spectrum of the irradiating terahertz pulse. (b) Schematic of the experimental setup.

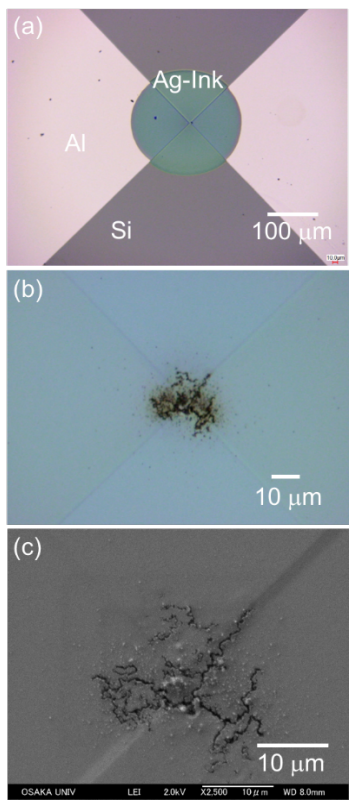

FIG. 2. Microscopic photographs of the sample (a) before and (b) after the terahertz pulse irradiation. (c) Scanning electron microscopy image recorded after the terahertz pulse irradiation.

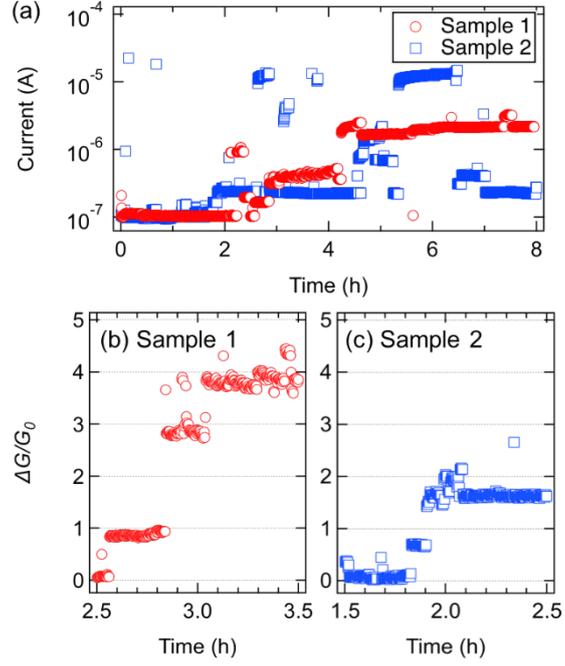

FIG. 3. (a) Measured currents under the terahertz pulse irradiation. Conductance changes normalized to the conductance quantum $G_0$ for the samples (b) 1 and (c) 2.

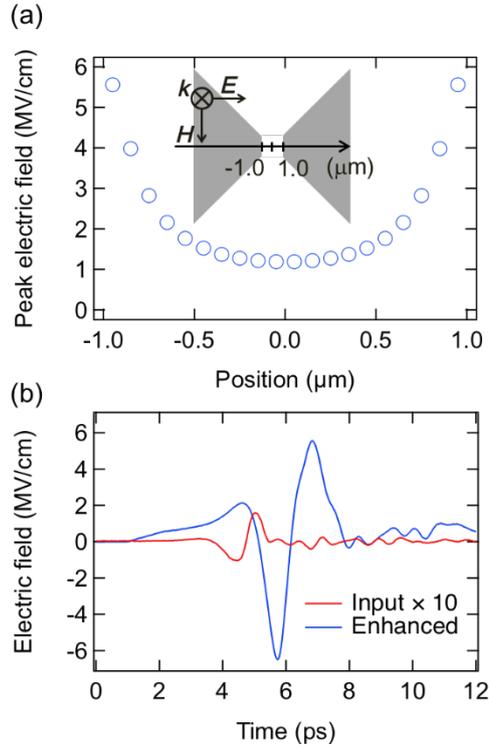

FIG. 4. (a) Simulated electric-field amplitude along the gap of the bowtie antenna corresponding to the peak amplitude of the terahertz pulse irradiation. (b) Time-domain waveform of the electric field at the maximum-field point of the gap.

[26] G. Antczak and G. Ehrlich, *Surface Diffusion* (Cambridge University Press, Cambridge, 2010).

**Supplementary material:**

**Quantized conductance observed during sintering of silver nanoparticles by intense terahertz pulses**

Keisuke Takano,[1,2]* Hirofumi Harada,[2] Masashi Yoshimura,[2] and Makoto Nakajima[2]

*[1]Center for Energy and Environmental Science, Shinshu University, 3–1–1 Asahi, Matsumoto, Nagano 390–8621, Japan.*
*[2]Institute of Laser Engineering, Osaka University, 2–6 Yamadaoka, Suita, Osaka 565–0871, Japan*

*\*Corresponding author: ksk_takano@shinshu-u.ac.jp*

**Sintering of silver nanoparticles on a dipole antenna array**

A dipole antenna array was fabricated on a high-resistivity Si substrate using photolithography. The unit cell of the antenna array consisted of a dipole pair with a 2-μm gap, as shown in Figs. S1(a) and S1(b). Silver nanometal ink (Ag1teH, nano-metal ink, ULVAC, Inc.) was spin-coated on the aluminum dipole antenna array, as shown in Fig. S1(c).

Irradiation with terahertz pulses (Fig. 1 in the main text) was performed for 1 h. After the irradiation, the color of the nanoparticles changed to black around the gap and edge of the dipole antenna pairs, as shown in Fig. S2. Figures S2(b) and S2(c) show scanning electron microscopy images of the blacked area, where cracks can be observed. In the cracked area, aggregation of nanoparticles is observed. The cracks indicate that the dispersant materials of the nanoparticles are peeled off; the volume decreased owing to the terahertz irradiation.

The aggregation of the nanoparticles leads to conduction at the antenna's gap. Figure S2(d) shows the transmission spectra of the sample before and after the terahertz pulse irradiation. The resonant frequency shifted from 0.53 THz to 0.31 THz. The conduction at the gap electrically connected the antenna pair. A longer antenna shows resonance at a lower frequency. The change of the resonant frequency is reproduced by finite-difference time-domain (FDTD) simulations, represented with the dashed curves in Fig. S2(d). In the simulation, the aluminum electrodes were assumed to be a perfect electric conductor and the resistivity between the antenna pair was assumed to be 10 kΩ. The nanoparticles were insulating owing to the dispersant material. Using the FDTD simulations, the electric field between the antenna pair was estimated to be larger than 5 MV/cm (Fig. S3). The gap resistance was assumed to be 100 Ω; the resonant dip at 0.31 THz was reproduced. The terahertz pulse irradiation led to conduction at the gap, and consequently the resonant frequency shifted from 0.53 THz to 0.31 THz.

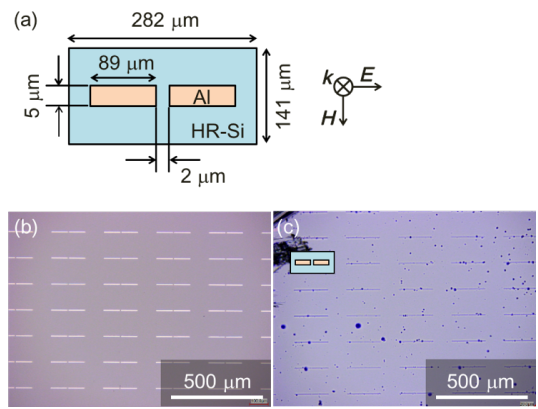

FIG. S1. (a) Schematic of the dipole antenna array. Micrographs of the dipole antenna array (a) before and (b) after the spin-coating of the silver nanometal ink.

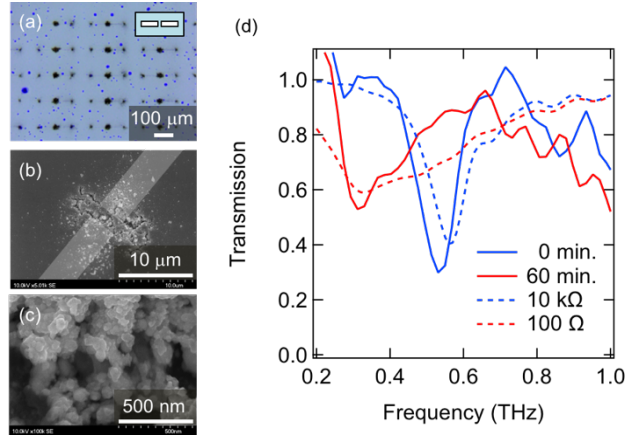

FIG. S2. (a) Micrograph of the sample after the 1-h-long irradiation with the terahertz pulses. The inset illustrates the unit cell position of the dipole pair array. (b, c) Scanning electron microscopy images of the blacked area. The white shadow indicates the position of the antenna, added as a guidance for a better visualization. (d) Transmission spectra before (blue solid) and after (red solid) the terahertz irradiation. The dashed curves represent the simulated transmission spectra with gap resistances of 10 kΩ (blue dashed) and 100 Ω (red dashed).

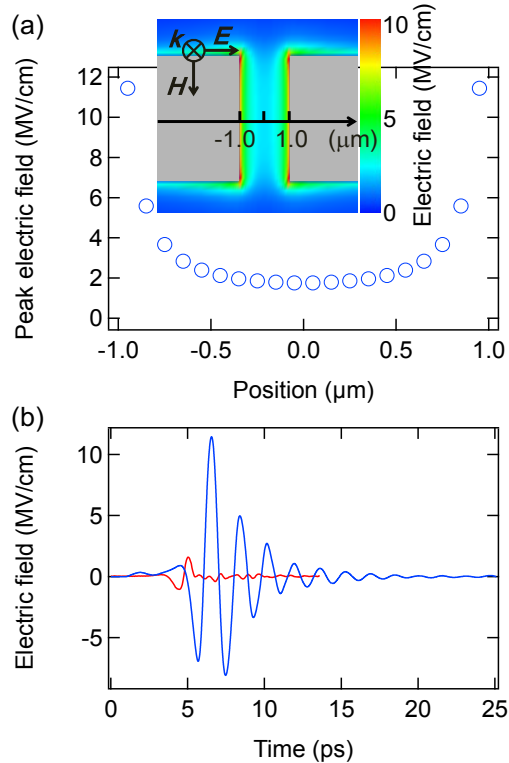

FIG. S3. Simulated electric field amplitude along the gap of the dipole antenna pair corresponding to the peak amplitude of the terahertz pulse irradiation. (b) Time-domain waveform of the electric field at the maximum-field point of the gap.

**Laser irradiation on silver nanoparticles**

The average power of the terahertz pulses was 1.0 mW (measured with THz-5B-MT, Gentec-EO). The repetition rate and beam diameter were 1 kHz and approximately 500 μm, respectively. The terahertz pulses caused color change of the nanometal ink. For comparison, we tested near-infrared laser pulse irradiation with the same average power. The bowtie antenna gap with the nanometal ink was irradiated with regenerative amplified laser pulses with a central wavelength of 800 nm, pulse width of 50 fs, repetition rate of 1 kHz, beam diameter of 500 μm, and average power of 1 mW. In this experiment, the bowtie antenna was fabricated on a fused silica substrate to avoid photocarrier excitation. Although the irradiation continued for 10 h, no significant change was observed, as shown in Fig. S4.

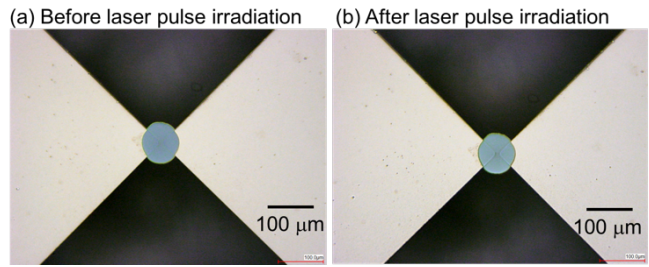

FIG. S4. Micrographs of the bowtie antenna with silver nanoparticles (a) before and (b) after the near-infrared laser pulse irradiation.